\documentclass[preprint,floats,aps,epsfig,notfootinbib,amssymb]{revtex4}
\usepackage{subfigure}
\usepackage{graphics,psfrag}
\usepackage{cancel}
\def\gev{\rm GeV}

\def\ttb{t\bar t}
\def\etmiss{E\!\!\!\!\slash_{T}}
\def\ptmiss{p\!\!\!\slash_{T}}
\def\pslash{\not{\hbox{\kern-4pt $p$}}}
\def\qslash{\not{\hbox{\kern-4pt $q$}}}
\def\lv{\not{\hbox{\kern-4pt $L$}}}
\def\lsim{\mathrel{\raise.3ex\hbox{$<$\kern-.75em\lower1ex\hbox{$\sim$}}}}
\def\gsim{\mathrel{\raise.3ex\hbox{$>$\kern-.75em\lower1ex\hbox{$\sim$}}}}
\def\ifmath#1{\relax\ifmmode #1\else $#1$\fi}
\def\beq{\begin{equation}}
\def\eeq{\end{equation}}
\def\bea{\begin{eqnarray}}
\def\eea{\end{eqnarray}}
\def\nn{\nonumber}

\usepackage{graphicx}
\usepackage{bm}
\begin{document}
\draft

\preprint{
   {\vbox {
      \hbox{\bf FERMILAB-PUB-08--098-T}
      \hbox{\bf MADPH-08-1511}	
      \hbox{\bf NSF-KITP-08-64}
      \hbox{arXiv: 0806.3966 [hep-ph]}
      }}}
\vspace*{2cm}

%
\title{Heavy Quarks Above the Top at the Tevatron}
\vspace*{0.25in}   
\author{Anupama Atre$^{1,2}$, Marcela Carena$^{1,2,3}$, Tao Han$^{2,4}$ and Jos\'e
Santiago$^{2,5}$
\footnote{avatre@fnal.gov\\ carena@fnal.gov\\ than@hep.wisc.edu\\ santiago@itp.phys.etzh.ch.}} 
\affiliation{\vspace*{0.1in}
$^1$ Fermi National Accelerator Laboratory, MS106, P.O.Box 500, IL 60510, U.S.A.\\
$^2$ Kavli Institute of Theoretical Physics, University of California, Santa Barbara, CA 93107, U.S.A.\\
$^3$ Department of Physics and Enrico Fermi Institute, University of Chicago, Chicago, IL 60637, U.S.A.\\
$^4$ Department of Physics, University of Wisconsin, 1150 University
Avenue, Madison, WI 53706, U.S.A.\\
$^5$ Institute for Theoretical Physics, ETH, CH-8093, Z\"{u}rich, Switzerland.
\vspace*{0.25 in}} 

\begin{abstract}
Recent developments in models with warped extra dimensions have opened
new possibilities for vector-like quark studies at hadron colliders. These new vector-like quarks can mix sizably with light Standard Model quarks without violating low energy constraints. We perform a model-independent analysis to determine the Tevatron reach in the search for new quarks. We find that the Tevatron has great potential to observe such quarks via their electroweak single production due to their mixing with valence quarks. With 4 (8) fb$^{-1}$ integrated luminosity, one may reach a  5$\sigma$ statistical significance for a heavy quark of mass 580 (630) GeV if the heavy quark-Standard Model quark mixing parameter is order one. 
\end{abstract} 

\maketitle

\section{Introduction}

The discovery of the top quark at the Fermilab Tevatron completed the three generations of fermions as the fundamental structure of matter fields in the  Standard Model (SM). With the large data sample being accumulated, the CDF and D0 experiments at the Tevatron are in a good position to search for heavier states at the high energy frontier. New vector-like quarks with sizable couplings to the SM quarks are a well-motivated extension of the SM, as they naturally appear in many theories beyond the SM. 

Due to the precision with which the couplings of light quarks have been measured, new vector-like quarks are typically allowed to mix sizably only with the third generation, mainly with the top quark. However, there can be cases in which corrections to the couplings of the SM quarks due to their mixing with heavy quarks can cancel,  leaving no observable trace of the existence of heavy quarks in SM interactions \cite{del Aguila:2000rc}. The simplest possibility, that we discuss in detail in the Appendix, is to introduce two degenerate doublets, with hypercharges $7/6$ and $1/6$, that only have Yukawa mixing with $u_R$
\cite{Carena:2006bn}, in the basis of diagonal Yukawa couplings in the up-type quark sector. This also ensures that flavor constraints are satisfied. Such scenarios can occur naturally in models with warped extra dimensions {\it with custodial protection of the $Zbb$ coupling} \cite{Agashe:2006at}.  The relevant part of the Lagrangian reads,
\bea
\nn
\mathcal{L} = \mathcal{L}_\mathrm{K}
&-&\Big[ \lambda_u \bar{q}^{(0)}_L \tilde{\varphi} u^{(0)}_R
+\lambda^i_d V_{ui} \bar{q}^{(0)}_L \varphi d^{(0)i}_R \\
&+&\ \ \lambda_Q \big(\bar{Q}^{(0)}_L \tilde{\varphi}+\bar{X}^{(0)}_L \varphi\big)
u^{(0)}_R
+m_Q\big(\bar{Q}_L^{(0)} Q_R^{(0)} + \bar{X}_L^{(0)} X_R^{(0)}\big)
+ \mathrm{h.c.} \Big],
\eea
where we have only explicitly written the up quark for the SM sector with electric charge $2/3$, $\mathcal{L}_\mathrm{K}=\bar{\psi} i \cancel{D} \psi$ is the sum of the diagonal kinetic terms (with covariant derivatives, thus including gauge couplings) for all the fields in the theory, $i=1,2,3$ are family indices, $V_{ui}$ is the first line of a unitary matrix (the CKM matrix in the absence of new physics), $\varphi$ is the SM Higgs field and $\tilde{\varphi}=i \sigma^2 \varphi^\ast $. The superscript $(0)$ denotes that the fields are not mass eigenstates and $X_{L,R}^{(0)}$ and $Q_{L,R}^{(0)}$ are the two new vector-like doublets with hypercharges $7/6$ and $1/6$, respectively. 

The couplings in the physical basis can be easily computed, as discussed in the Appendix. The result is that the corrections to the gauge couplings of the SM quarks are negligible. There are four new heavy quarks in the spectrum: one with electric charge $-1/3$, one with electric charge $5/3$ and two with electric charge $2/3$. The phenomenologically relevant couplings of the new heavy quarks with the SM quarks are listed in Table~\ref{tab:coup} in the limit $v \ll m_Q$, where $m_Q$ is the heavy quark mass, $v=174$ GeV is the SM Higgs vacuum expectation value, $g$ is the weak coupling constant and $c_W^{}$ is the cosine of the weak angle. All the other couplings of these new quarks to the SM quarks are extremely suppressed and therefore irrelevant.  More interestingly in the models with extra dimensions that motivated our study, $\lambda_Q$ can be naturally order one.  This large coupling, together with very distinctive kinematics, makes single production of a heavy quark an ideal process for its discovery. 

\begin{table}[tb]
{
\begin{tabular}{c |c | c | c | c }
\hline
State & $q^-$ & $q^+$  & $q^d$ & $\chi^u$ \\
\hline
Electric Charge & $2/3$ & $2/3$  & $-1/3$ & $5/3$ \\
\hline
\multicolumn{5}{c}{Coupling to $u_R$}\\
\hline
CC &  &   & $\frac{-g}{\sqrt 2} \frac{v}{m_Q} \lambda_Q$ & $\frac{-g}{\sqrt 2} \frac{v}{m_Q} \lambda_Q$ \\
NC & $\frac{-g}{\sqrt 2 c_W} \frac{v}{m_Q} \lambda_Q$ &   & & \\
Yukawa & &  $\sqrt{2} \lambda_Q$ & &\\
\hline
\end{tabular}
}
\caption[]{Couplings of new heavy quarks to SM up-type quark in the limit $v \ll m_Q$, where $m_Q$ is the heavy quark mass, $v=174$ GeV is the SM Higgs vacuum expectation value, $g$ is the weak coupling constant and $c_W^{}$ is the cosine of the weak angle.}
\label{tab:coup}
\end{table}

Motivated by the above set up we investigate the potential of the Tevatron to find new quarks and perform a model-independent analysis as described below. Let us consider two new quarks, $U$ and $D$, with masses $m_{U,D}$ and electric charges $Q_U=2/3$ and $Q_D=-1/3$, respectively. Based on the discussion above, we assume they do not induce anomalous couplings among the SM quarks and they have the following charged current (CC) and neutral current (NC) gauge interactions to the first generation quarks,
\beq
\label{Coupling}
 \frac{g}{\sqrt{2}} W_\mu^+ 
(\kappa_{uD}\ \overline{u}_R \gamma^\mu D_R
+\kappa_{dU}\ \overline{d}_R \gamma^\mu U_R)
+ \frac{g}{2 c_W} Z_\mu(\kappa_{uU}\ \overline{u}_R \gamma^\mu U_R 
+ \kappa_{dD}\ \overline{d}_R \gamma^\mu D_R) + \mathrm{h.c.} \qquad
\eeq
The coupling strength is parameterized by $\kappa_{qQ}$ in a model-independent manner as 
\beq
\kappa_{qQ} = ({v}/{m_Q}) \tilde{\kappa}_{qQ},
\eeq
where the dimensionless parameter $\tilde{\kappa}_{qQ}$ encodes the model-dependence. Thus, the relevant couplings in the model we have discussed, in the limit, $m_Q \gg v$ are the ones in Eq.~(\ref{Coupling}), with 
$\tilde{\kappa}_{uU} \approx -\sqrt{2}\lambda_Q$ and 
$\tilde{\kappa}_{uD} \approx -\lambda_Q$. Note, however, that our parameterization is completely model-independent and it still includes the case of lighter $m_Q$. A similar model with two doublets of hypercharges $1/6$ and $-5/6$ that mix only with the $d_R$ quark will generate the other two couplings in Eq.~(\ref{Coupling}). We have not explicitly written down the heavy quark Higgs couplings because they do not contribute  appreciably to the production process of our interest. For the purpose of decay properties of heavy quarks the coupling of fermions with Higgs can be reabsorbed in the definition of decay branching ratios that we will leave as a free parameter in our analysis. Also, the extra quarks with exotic charges ($5/3$ or $-4/3$) can be trivially included by multiplying the corresponding production cross section by the appropriate number of quarks. In Eq.~(\ref{Coupling}) we have only written down right-handed (RH) couplings, which appear in the case of vector-like doublets. Left-handed (LH) couplings will appear in the case of vector-like singlets. Since we will not make use of angular correlations, our results do not depend on the choice of the chiral couplings appreciably. 
\begin{figure}[t,b]
{\includegraphics[width=0.465\textwidth,clip=true]{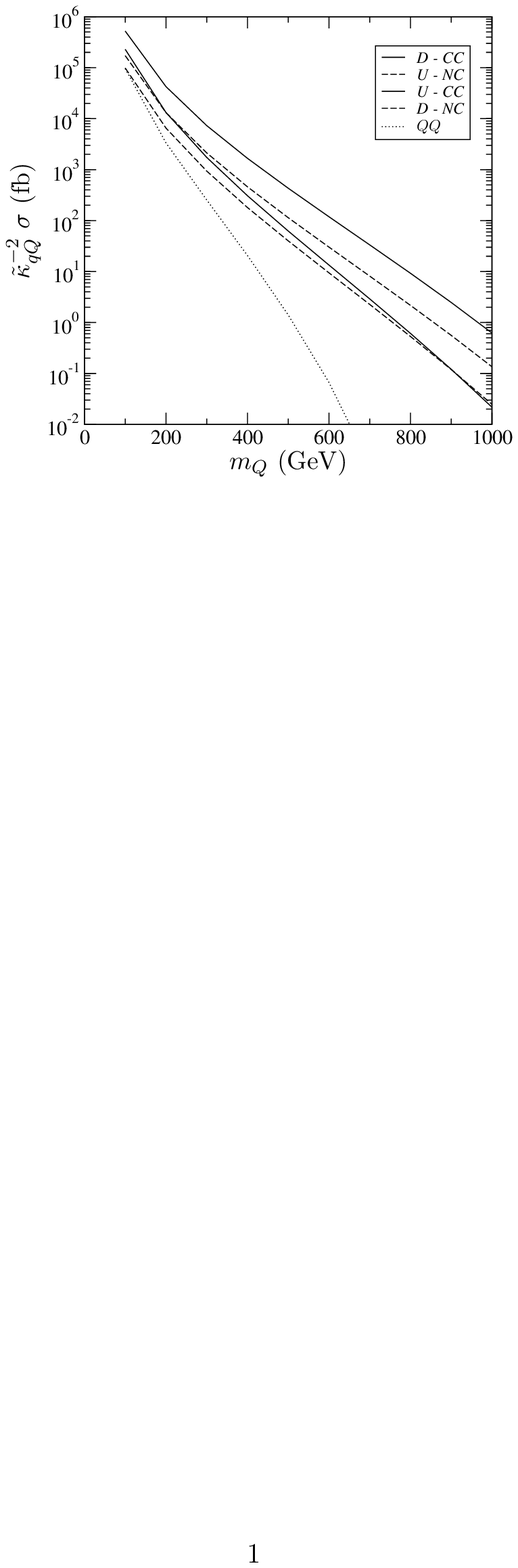}}
\caption{ 
Total cross sections for heavy quark production versus its mass $m_Q^{}$ at the Tevatron with $\sqrt s=1.96$ TeV in $p\bar p$ collisions. Solid curves are for single production via charged current (CC) of $D+\overline D$ (upper) and 
$U+\overline U$ (lower); dashed  curves are for single production via neutral current (NC) of $U+\overline U$ (upper) and $D+\overline D$ (lower); and the dotted curve is for pair production from QCD.}
\label{fig:Qs}
\end{figure}

\section{Heavy quark production}

Heavy quarks can be produced in pairs via strong QCD interactions
\beq
q\bar q,\ gg \to Q\bar Q.  
\eeq
Alternatively, a heavy quark can also be produced singly via the weak interactions as in Eq.~(\ref{Coupling})
\beq
q q'  \stackrel {V^*} {\longrightarrow} q_1 Q,
\eeq
where $V = W$ or $Z$ gauge boson. The production cross sections for these processes are shown in Fig.~\ref{fig:Qs} at the Tevatron energy 
($\sqrt s=1.96$ TeV) versus the heavy quark mass $m_Q$, where the NLO corrections to the total cross section with respect to our tree-level results (the 
$K$ factor) have been taken into account as $K \approx 1.5$ for pair production \cite{QCDcorrect}, and $K\approx 0.96$ for single production \cite{QCDcorr}. The pair production (dotted curve) is completely dominated by valence quark annihilation. The current bound from direct searches at the Tevatron experiments is $m_Q > 284\ (270)$ GeV at $95\%$ C.L. for heavy up (down) type quarks decaying via CC (NC) \cite{mqbound}. This is unlikely to improve dramatically as the cross section falls off sharply due to phase space suppression and decreasing parton luminosity at large $x$ values.

On the other hand, single heavy quark production has the advantage of less phase space suppression and longitudinal gauge boson enhancement of order 
$m_Q^2/M_V^2$ at higher energies. Due to the participation of $u,d$ valence quarks in the initial state with the coupling strength given in Eq.~(\ref{Coupling}), the cross section can be substantial and it falls more slowly for a higher mass. For a model-independent presentation, the coupling parameters, generically denoted by $\tilde{\kappa}$, have been factored out. The full leading-order
matrix elements for $q' q \to q_1 Q \to q_1 q_2 f \overline f$ with $qW^*$ and $qZ^*$ fusion have been calculated using helicity amplitudes and cross-checked against other available packages. For instance, for a mass as heavy as 600 GeV, with $\tilde\kappa\sim 1$, the cross section is of the order of 100 fb for each new quark. Their relative sizes are determined by the corresponding valence quark density in the initial state. In our analysis we use CTEQ6L1 parton distribution functions \cite{Pumplin:2002vw} and choose the factorization scale, 
$\mu_F = M_W, M_Z$ for the signal \cite{Han:1992hr}. For the background, we choose the factorization and renormalization scales to be 
$\mu_F = \mu_R =  \sqrt{\hat s}/2$.

\section{Heavy quark decay}

The singly produced heavy quarks will decay into jets and gauge or Higgs bosons through gauge and Yukawa interactions. The allowed channels are
\bea
D \to W^- u,\ Z d,\ h d,\quad  U \to W^+ d,\ Z u,\ h u.
\eea
For the remainder of this work we will concentrate on the gauge boson channels. To perform a model-independent study, we parameterize the cross section under the narrow-width approximation as
\beq
\sigma(pp \rightarrow q_1 q_2 f \overline f) \equiv S^{\scriptscriptstyle CC(NC)}_Q\ \sigma^{\scriptscriptstyle CC(NC)}_{prodn}\ Br(V\to f\bar f),
\eeq
 where $\sigma^{\scriptscriptstyle CC(NC)}_{prodn}$ is only dependent on the c.m. energy and mass of the heavy quark and 
 $S^{\scriptscriptstyle CC(NC)}_Q$ encode the model-dependent parameters and are defined as
\bea
\nn
S_{\scriptscriptstyle D}^{\scriptscriptstyle CC} &\equiv& (\tilde\kappa^{2}_{uD} + \alpha^{\scriptscriptstyle CC}_{\scriptscriptstyle D}\ \tilde\kappa^{2}_{dD})\ Br(D \rightarrow q W), \\
\nn
S_{\scriptscriptstyle U}^{\scriptscriptstyle CC} &\equiv& (\tilde\kappa^{2}_{dU} + \alpha^{\scriptscriptstyle CC}_{\scriptscriptstyle U}\ \tilde\kappa^{2}_{uU})\ Br(U \rightarrow q W), \\
\nn
S_{\scriptscriptstyle D}^{\scriptscriptstyle NC} &\equiv& (\tilde\kappa^{2}_{dD} + \alpha^{\scriptscriptstyle NC}_{\scriptscriptstyle D}\ \tilde\kappa^{2}_{uD})\ Br(D \rightarrow q Z), \\
S_{\scriptscriptstyle U}^{\scriptscriptstyle NC} &\equiv& (\tilde\kappa^{2}_{uU} + \alpha^{\scriptscriptstyle NC}_{\scriptscriptstyle U}\ \tilde\kappa^{2}_{dU})\ Br(U \rightarrow q Z),
\eea
where $\alpha^{\scriptscriptstyle CC}_{\scriptscriptstyle Q } \equiv \sigma^{\scriptscriptstyle NC}_{prodn}/\sigma^{\scriptscriptstyle CC}_{prodn}$ and 
$\alpha^{\scriptscriptstyle NC}_{\scriptscriptstyle Q } \equiv \sigma^{\scriptscriptstyle CC}_{prodn}/\sigma^{\scriptscriptstyle NC}_{prodn}$ are the ratios of the production cross section of the heavy quark via CC and NC and can be deduced from Fig.~\ref{fig:Qs}. In the case of degenerate bidoublets, only one gauge boson decay mode is available for each new quark and  
$Br[Q \rightarrow q W(Z)]$ is $100\%$. For instance, if 
$Br[D \rightarrow u W] = Br[U \rightarrow u Z] = 100\%$, then
$S_{\scriptscriptstyle D}^{\scriptscriptstyle CC} = \tilde\kappa^{2}_{uD}$ and $S_{\scriptscriptstyle U}^{\scriptscriptstyle NC} = \tilde\kappa^{2}_{uU}$.

\renewcommand{\arraystretch}{1.4}
\begin{table}[t]
\begin{center}
\begin{tabular}{|l|c|c|c|} \hline
channels & Basic cuts (\ref{eq:basiccuts}) & High $p_T$ (\ref{eq:highptcuts}) & $m_Q$ (\ref{eq:masscuts})  \\
\hline
$D\to W^\pm q$ & 270 & 190 & 160 \\
$U\to W^\pm q$ & 49 & 35 & 29 \\
\hline
$W^\pm + 2j$  & 79000 & 1200 & 280 \\
$W^\pm W^\mp  (\to 2j)$  & 1500 & 15 & 1.4\\
$W^\pm Z(\to 2j)$  & 230 & 4.7 & 0.52  \\
single top: $W^\pm b\ j$  & 330 & 10 & 2.9 \\
$\ttb$: fully leptonic  & 170 (79) & 2.0 & 0.40 \\
$\ttb$: semi-leptonic  & 600 & 0.19 & - \\
\hline
\end{tabular}
\caption{Total cross-sections (in fb) for the signal with $m_Q=400$ GeV and $S^{\scriptscriptstyle CC}_{\scriptscriptstyle Q} = 1$ and  the leading SM backgrounds at the Tevatron before and after the kinematical cuts in steps described in the text.  $D+\overline D$ and $U+\overline U$ and the leptons 
$\ell = e, \mu$ have been counted for. For $\ttb$, the numbers in parentheses in the second column include a veto on events with two isolated leptons.}
\label{tab:cc}
\end{center}
\end{table}

\renewcommand{\arraystretch}{1.4}
\begin{table}[t]
\begin{center}
\begin{tabular}{|l|c|c|c|} \hline
channels & Basic cuts (\ref{eq:basiccuts}) & High $p_T$ (\ref{eq:highptcuts}) & $m_Q$ (\ref{eq:masscuts})  \\
\hline
$D\to Z(\rightarrow \ell \ell) q$ & 8.8 & 6.0 & 5.7 \\
$U\to Z(\rightarrow \ell \ell) q$ & 22 & 15 & 15 \\
\hline
$Z(\rightarrow \ell \ell) + 2j$  & 7000 & 120 & 14 \\
$Z(\rightarrow \ell \ell)W^\pm(\to 2j)$  & 60 & 0.65 & 0.08 \\
$Z(\rightarrow \ell \ell)Z(\to 2j)$  & 55 & 1.1 & 0.11 \\
$\ttb$: fully leptonic  & 160 (1.7) & - & - \\
\hline
\end{tabular}
\caption{Same as in Table~\ref{tab:cc} but with 
$S^{\scriptscriptstyle NC}_{\scriptscriptstyle Q} = 1$. The numbers in parentheses in the second column include a veto on events with missing energy. See text for details.}
\label{tab:nc}
\end{center}
\end{table}

\renewcommand{\arraystretch}{1.4}
\begin{table}[t]
\begin{center}
\begin{tabular}{|l|c|c|c|} \hline
channels & Basic cuts (\ref{eq:basiccuts}) & High $p_T$ (\ref{eq:highptcuts}) & $m_Q$ (\ref{eq:masscuts})  \\
\hline
$D\to Z(\rightarrow \nu \nu) q$ & 31 & 22 & 18  \\
$U\to Z(\rightarrow \nu \nu) q$ & 79 & 56 & 46   \\
\hline
$Z(\rightarrow \nu \nu) + 2j$  & 28000 & 630 & 160 \\
$Z(\rightarrow \nu \nu) W^\pm(\to 2j)$  & 240 & 3.4 & 0.30 \\
$Z(\rightarrow \nu \nu) Z(\to 2j)$  & 220 & 6.1 & 0.76 \\
$\ttb$: fully leptonic  & 260 (12) & 1.5 & 0.89 \\
$\ttb$: semi-leptonic  & 880 (290) & 2.3 & 1.1 \\
\hline
\end{tabular}
\caption{Same as in Table~\ref{tab:nc} but with $\ell = e, \mu$ and 
$\nu = \nu_e, \nu_\mu, \nu_\tau$. The numbers in parentheses in the second column include a veto on events with isolated leptons.}
\label{tab:ncnu}
\end{center}
\end{table}

\section{Observability of the heavy quark signal}

For the signal identification, we also require the clean leptonic decay modes of the gauge boson with $\ell=e,\mu$ and $\nu = \nu_e, \nu_\mu, \nu_\tau$. Although the inclusion of $\tau$ lepton in the final state could increase the signal statistics, for simplicity we ignore this experimentally more challenging channel. Focusing on the single heavy quark production, the three channels of the final states under consideration are
\beq
\ell^\pm \etmiss\ 2j,\quad \ell^+\ell^-\ 2j,\quad \etmiss\ 2j,
\eeq
from $Q$ decaying to a $W(\to \ell^\pm \nu)$ and $Z (\rightarrow \ell^+ \ell^-,\  \nu \bar{\nu})$, respectively. We perform a partonic analysis and simulate the detector resolution by smearing the energies of the leptons and jets by the Gaussian form according to 
${\Delta E_\ell}/{E_\ell} = {0.135}/{\sqrt{E_\ell/\mathrm{GeV}}} \oplus 0.02$ and 
${\Delta E_j}/{E_j} = {0.75}/{\sqrt{E_j/\mathrm{GeV}}} \oplus 0.03$ respectively \cite{Quadt:2007jk}. We select the events to contain isolated leptons/large 
$\etmiss$ and two jets with the following basic acceptance cuts \cite{d0}
\bea
\nn
&& p_T(\ell),\  p_T(j),\  \etmiss > 15\ {\gev},\ \  |\eta_\ell | < 2,\ |\eta_j| < 3, \\
&& \Delta R(jj) > 0.7,\ \Delta R(j\ell) > 0.5,\ \Delta R(\ell\ell) > 0.3.
\label{eq:basiccuts}
\eea
To quantify the signal observability, we must consider the SM backgrounds. The
irreducible backgrounds are 
\begin{itemize}
\item $W+2$ jets, $Z+2$ jets with $W, Z$ leptonic decays;
\item $W^+W^-, W^\pm Z,$ and $ZZ$ with semi-leptonic decays;
\item single top production leading to $W^\pm b\ q$.
\end{itemize}
With the basic cuts in Eq.~(\ref{eq:basiccuts}), we list the cross sections of the signal with $m_Q=400$ GeV and 
$S^{\scriptscriptstyle CC(NC)}_{\scriptscriptstyle Q} = 1$ for different channels and the leading SM backgrounds in the second column in 
Tables \ref{tab:cc} $-$ \ref{tab:ncnu}. The background from $t \bar t$ is relevant. However, with the cuts as described in 
Eqs.~(\ref{eq:basiccuts}) $-$ (\ref{eq:masscuts}) and some additional cuts described below we can essentially eliminate the $\ttb$ background. For the case of CC decay modes of the signal, we can impose a veto on a second isolated lepton (defined as 
$p_T(\ell) > 15\ \mbox{GeV},\ |\eta_\ell| < 2$ and $\Delta R(j\ell) > 0.5$) to reduce the $\ttb$ background from the fully leptonic decay mode. For the NC decay mode of the signal with leptons in the final state, we can veto events with missing energy ($\etmiss > 15\ \mbox{GeV}$). For the NC decay mode into neutrinos, we can veto events with any isolated leptons. The cross-section with the lepton/missing energy vetoes for the $\ttb$ background are listed in parentheses in the second column of Tables \ref{tab:cc} $-$ \ref{tab:ncnu}. To further reduce $\ttb$ events, we can veto events with b-tags (this has virtually no effect on our signal) and require that the two leptons reconstruct the Z boson for the NC signal. The semi-leptonic decay mode can be reduced further by vetoing events where any two jets reconstruct a W-boson or if any three jets reconstruct the top quark. Note that even with just the cuts in 
Eqs.~(\ref{eq:basiccuts}) $-$ (\ref{eq:masscuts}) and the simple vetoes which do not affect our signal significantly, the $\ttb$ background is negligible compared to the leading background and even the signal. Hence we can justifiably neglect the $\ttb$ background in our analysis. 

\begin{figure}[tb]
{\includegraphics[width=7.6cm,clip=true]{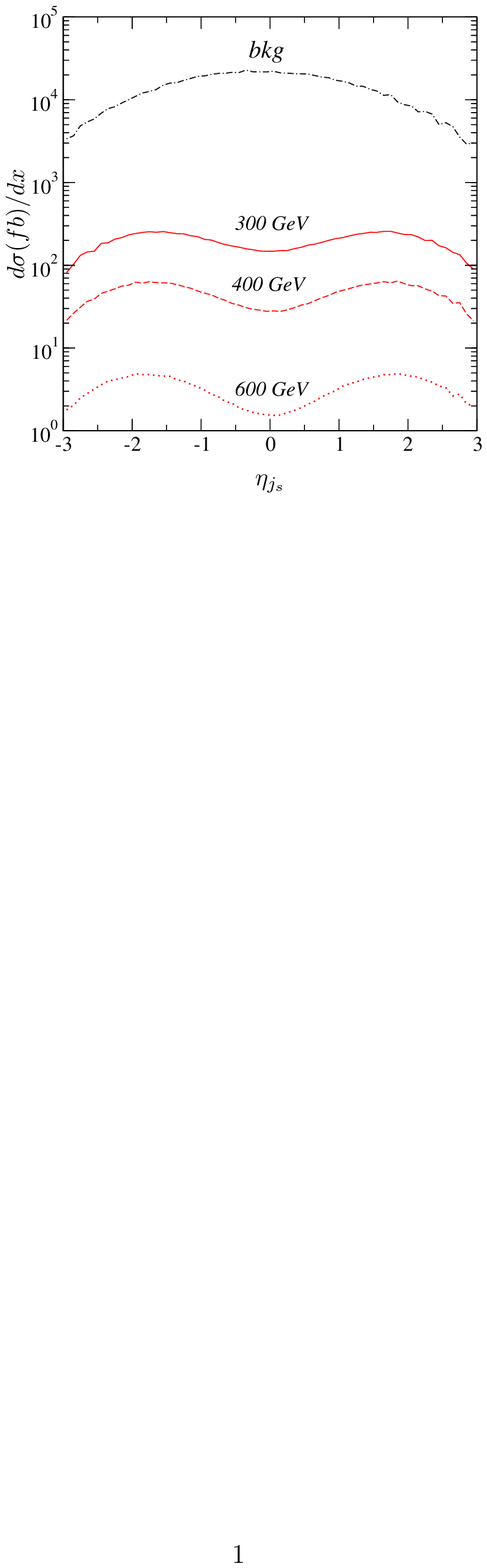}
\vspace {7.5 mm}
\includegraphics[width=7.75cm,clip=true]{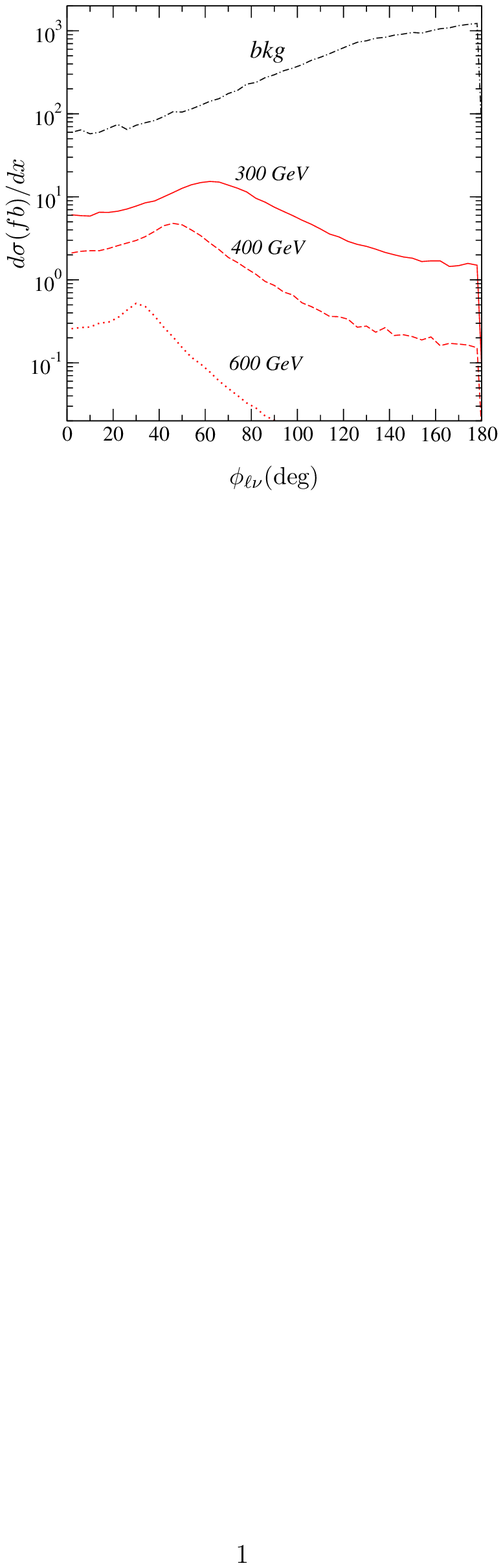}
\includegraphics[width=7.75cm,clip=true]{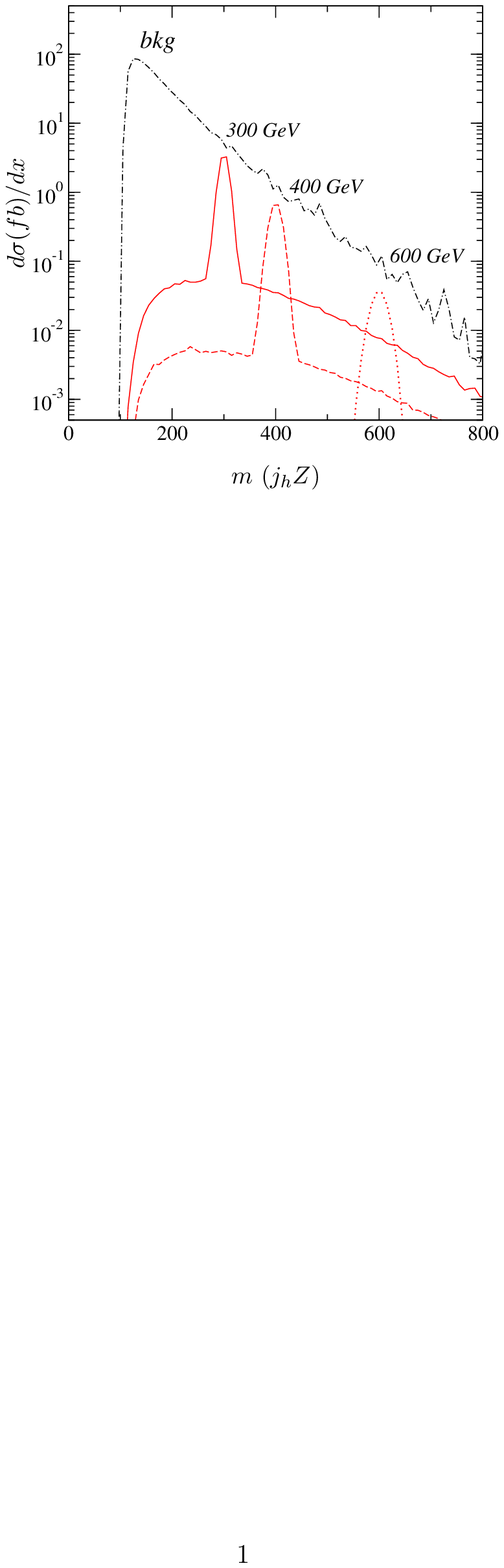}
\includegraphics[width=7.75cm,clip=true]{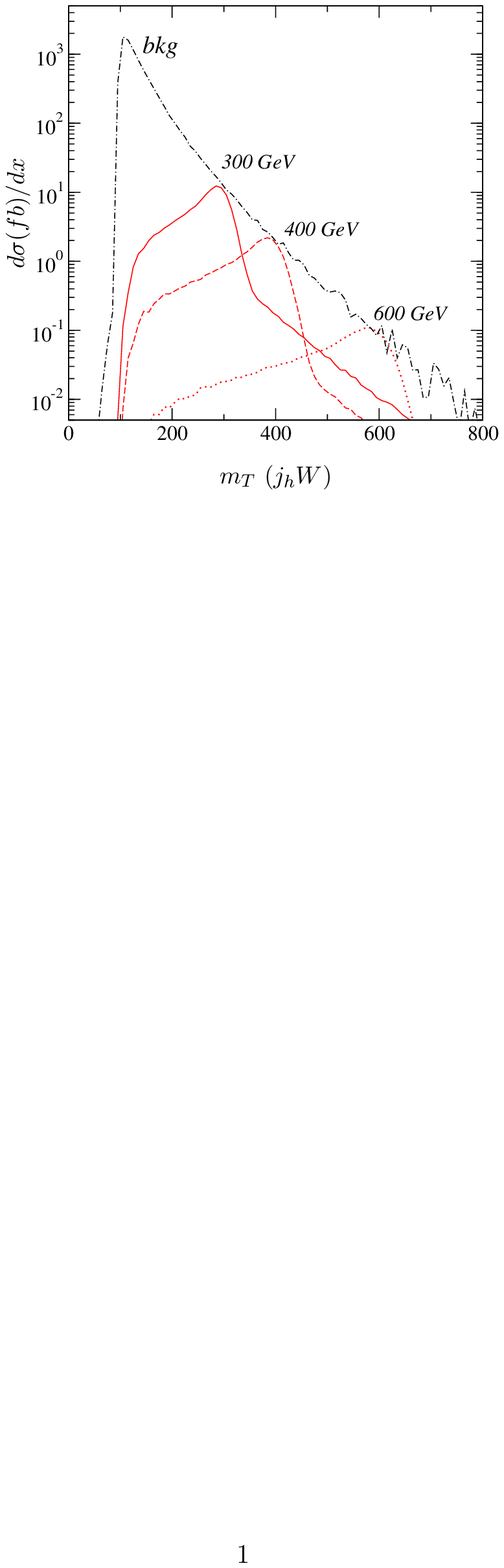}}
\caption{ 
(a) Top left: pseudo-rapidity distribution for the soft jet. (b) top right: azimuthal angle between $p_T(\ell)$ and $\ptmiss$ (c) bottom left: invariant mass distribution for the reconstructed heavy quark from visible $Z$ decay (d) bottom right: transverse mass distributions for the reconstructed heavy quark.} 
\label{fig:dist}
\end{figure}

While the background cross sections can be large to begin with, it is important to notice the qualitative difference of the kinematics between the signal and backgrounds. First, one of the two jets is associated with $W,Z$ $t$-channel exchange with a typical transverse momentum $\sim M_W/2$. More importantly, the second jet is from the heavy quark decay that makes it very energetic with a Jacobian peak near $p_T(j)\approx  {1\over 2} m_Q (1-M_W^2/m_Q^2)^{1/2}$. Using the $p_T$ of the jets as a discriminant gives very good accuracy in
identifying the correct jets, especially for high masses. Hence we identify the hardest jet ($j_h$) as the one from heavy quark decay and the softer jet ($j_s$) as the one associated with $W/Z$ $t$-channel exchange. Similarly, the $W/Z$ from the heavy quark decays are also very energetic. We can thus design a larger $p_T$ cut on the hard jet and the reconstructed $W/Z$ to separate the signal from the background. Next, we note that the pseudo-rapidity of the soft jet
associated with $W/Z$ $t$-channel exchange peaks at $|\eta_{j_s}| \sim 2$, leading to the forward jet tagging to enhance the signal over the backgrounds. The pseudo-rapidity distribution for the soft jet of both signal and backgrounds are shown in Fig.~\ref{fig:dist}(a). The $W$ ($Z$) gauge boson from the heavy quark decay via the CC (NC) can be energetic depending on $m_Q$, leading to rather collimated final state leptons; while those leptons in the background still tend to be back-to-back. This is shown in Fig.~\ref{fig:dist}(b) for the azimuthal
angle separation between the leptons ($p_T(\ell)$ and $\ptmiss$) for CC. A similar distribution between $\ell^+ \ell^-$ is present for the NC channel with visible $Z$ decays. Also, the di-jet separation for the signal is large, typically back-to-back; while that for the QCD background tends to present a collinear singularity. In order to take advantage of these kinematical features we impose
\bea
\nn
& &p_T(j_h) > {1 \over 4}m_Q,\ \ p_T(W/Z) > {1\over 5} m_Q,\\
& &\Delta R(jj) > 1.5,\ \Delta R(j\ell) > 0.8,\ 0.5 < |\eta_{j_s}| < 3.0.
\label{eq:highptcuts}
\eea
In addition we use a mass based cut on the azimuthal angle of the leptons 
($\phi_{\ell \nu}, \phi_{\ell \ell})$ optimized for each mass. The results with these improved cuts are listed in the third column in 
Tables \ref{tab:cc} $-$ \ref{tab:ncnu}.   
Of most importance is the reconstruction of the mass peak for the resonant
particles. For a heavy quark decay with $Z\to \ell^+\ell^-$ in the final state, it is straightforward to form the invariant mass for the heavy quark $m_Q=M(Z, j_h)$, as shown in Fig.~\ref{fig:dist}(c) for the signal and backgrounds. For the final state with $W\to \ell\nu$ and $Z\to \nu\bar{\nu}$, one can define a  cluster transverse mass variable to be 
\bea
\nn
M_T^2 = \left(\sqrt{p_{TW,Z}^2+M_{W,Z}^2} + p_{Tj_h}\right)^2 -
\left(\vec p_{TW,Z} + \vec p_{Tj_h} \right)^2.
\eea
Those distributions are plotted in Fig.~\ref{fig:dist}(d). We suggest the following invariant mass cuts 
\bea
\nn
m_Q - {1 \over 4}m_Q &<& M_T(j_h W/Z) < m_Q + 50\ {\rm GeV},\\
m_Q - 30\ {\rm GeV} &<& M(j_h Z) < m_Q + 30\ {\rm GeV}.
\label{eq:masscuts}
\eea
The results with the above invariant mass cuts are listed in the last column in Tables \ref{tab:cc} $-$ \ref{tab:ncnu}. 

Several remarks are in order. Firstly, it is very interesting to notice the possibility of identifying the electromagnetic charge of the heavy quark produced. Note that the forward (backward) distribution of the soft jet should be correlated with the heavy anti-quark (quark) production. Moreover, this can be used as an indication for down-type or up-type heavy quark production by specifying the electromagnetic charge of the lepton. For e.g., an event with a backward soft jet and a positive (negative) lepton would indicate production of $U$ ($D$) heavy quark. Similarly, an event with a forward soft jet and a positive (negative) lepton would indicate production of $\overline D$ ($\overline U$) heavy quark. Secondly, in our analysis we included a single mass window cut based on an assumed mass for the heavy quark. This is standard to optimize the signal significance. However, an experimental analysis would include several mass window cuts and the appropriate statistical dilution factor. Thirdly, we have presented a partonic analysis with detector effects included through smearing. To simulate a realistic experimental environment, one would need to include showering and hadronization effects as well as real detector simulation. We have checked the credibility of our partonic simulation by including showering (ISR and FSR) and hadronization using PYTHIA \cite{Sjostrand:2006za} and realistic detector effects using PGS \cite{pgs} for $m_Q = 400$ GeV for the CC channel and the results are presented in Table \ref{tab:pgs}. In order to also
estimate the effects of some reducible backgrounds, we have analyzed W+jet events with higher jet multiplicity (vetoing events with  a third jet with $p_T>10$ GeV to reduce backgrounds). We have estimated that the effect of such multi-jet backgrounds reduces the $S/\sqrt{B}$ from the partonic analysis by $\lesssim 20\%$ for $m_Q=400$ GeV in the CC channel. However, we expect that the more refined optimization techniques used in experiments, which are beyond the scope of this study, will be able to offset the above effects. 

\renewcommand{\arraystretch}{1.4}
\begin{table}[t]
\begin{center}
\begin{tabular}{|l|c|c|c|c|} \hline
& $D\to W^\pm q$ & $W^\pm + 2j$   & $ \Bigl(\frac {S}{\sqrt B} \Bigr)_{fast\ sim}$ & $ \Bigl(\frac{S}{\sqrt B} \Bigr)_{partonic}$   \\
\hline
Basic cuts (\ref{eq:basiccuts})  & 200 & 28000 & 1.2 & 0.96 \\
High $p_T$ (\ref{eq:highptcuts}) & 120 & 390 & 6.1 & 5.5 \\
$m_Q$ (\ref{eq:masscuts})  & 84 &  90 & 8.9 & 9.6\\
\hline
\end{tabular}
\caption{Total cross-sections (in fb) for the signal with $m_Q=400$ GeV and $S^{\scriptscriptstyle CC}_{\scriptscriptstyle Q} = 1$ and  the leading SM background at the Tevatron including showering and hadronization using PYTHIA and detector simulation using PGS. The corresponding significance from the partonic simulation is listed in the last column for comparison.}
\label{tab:pgs}
\end{center}
\end{table}

\begin{figure}[tb]
{\includegraphics[width=0.46\textwidth,clip=true]{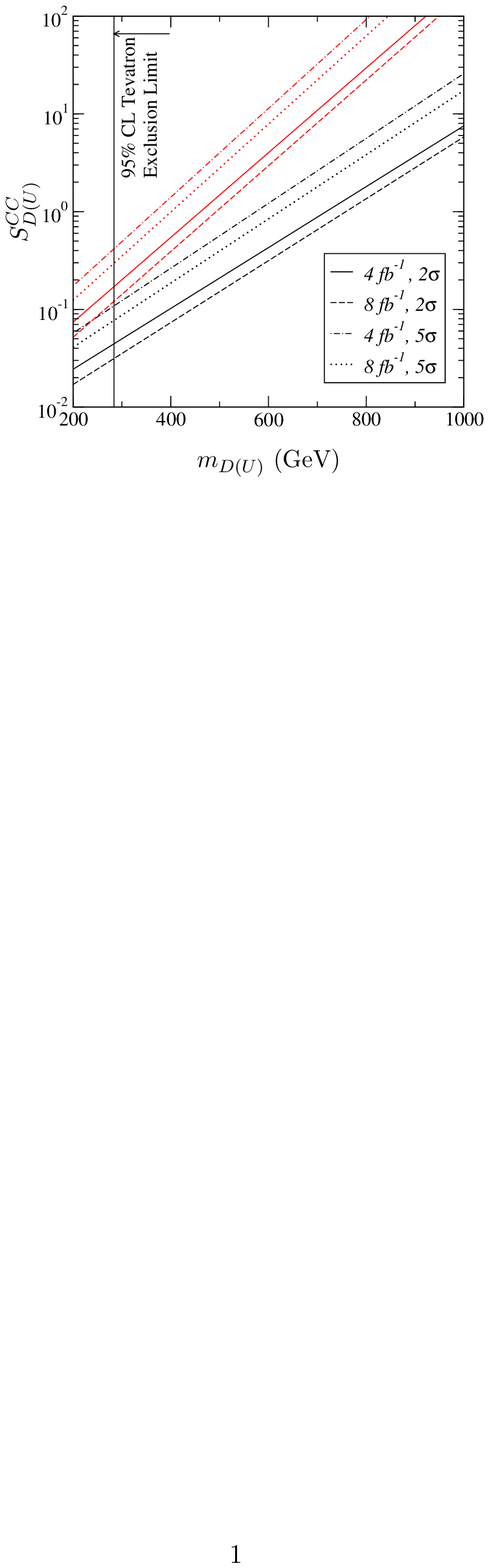}}
\caption{ 
Sensitivity plots in the plane of model-dependent parameter 
$S^{\scriptscriptstyle CC}_{\scriptscriptstyle Q}$ and heavy quark mass $m_Q$ for CC decay mode of heavy quark with 4 fb$^{-1}$ and 8 fb$^{-1}$ integrated luminosity. The top set (red) is for $m_U$ versus 
$S^{\scriptscriptstyle CC}_{\scriptscriptstyle U}$ and bottom set (black) is for $m_D$ versus $S^{\scriptscriptstyle CC}_{\scriptscriptstyle D}$.}
\label{fig:SenCC}
\end{figure}
\begin{figure}[tb]
\includegraphics[width=0.46\textwidth,clip=true]{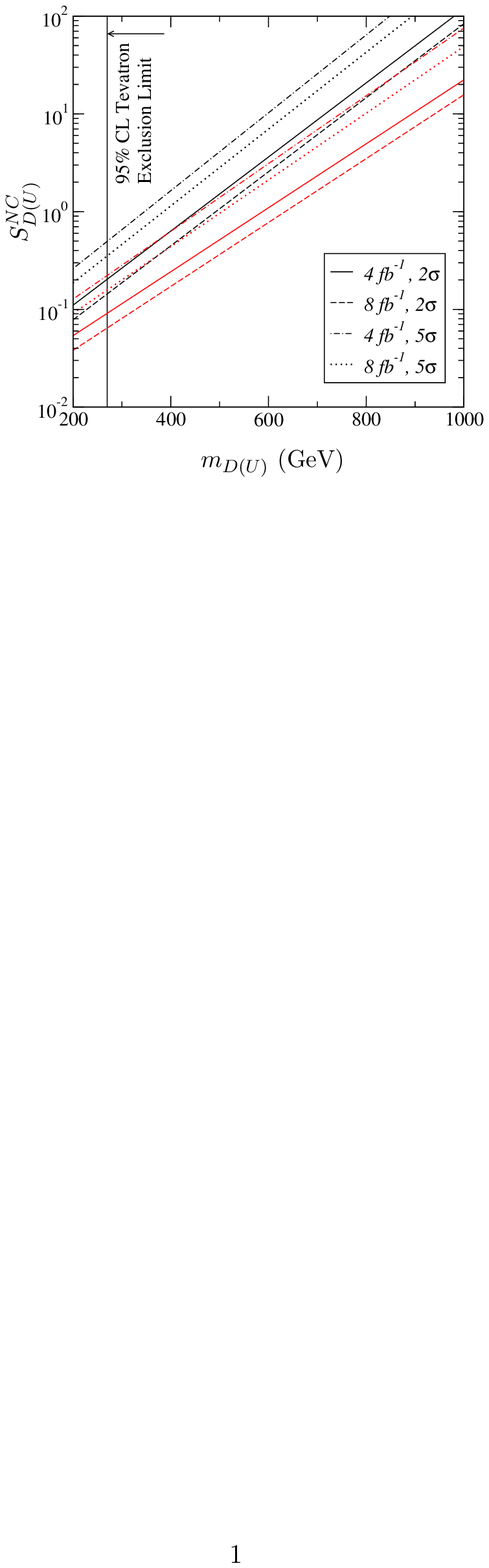}
\caption{ 
Sensitivity plots in the plane of model-dependent parameter 
$S^{\scriptscriptstyle NC}_{\scriptscriptstyle Q}$ and heavy quark mass $m_Q$ for NC decay mode of heavy quark with 4 fb$^{-1}$ and 8 fb$^{-1}$ integrated luminosity. The top set (black) is for $m_D$ versus 
$S^{\scriptscriptstyle NC}_{\scriptscriptstyle D}$ and bottom set (red) is for $m_U$ versus $S^{\scriptscriptstyle NC}_{\scriptscriptstyle U}$.}
\label{fig:SenNC}
\end{figure}

We estimated the statistical significance for a signal near the assumed mass peak and analyse the full parameter space in the coupling parameter 
$S_{\scriptscriptstyle Q}^{\scriptscriptstyle CC(NC)} $ and $m_Q$ plane. For an integrated luminosity of 4 and 8 fb$^{-1}$, we plot the 2$\sigma$ and 5$\sigma$ contours in Fig.~\ref{fig:SenCC} for the CC decay mode and in  Fig.~\ref{fig:SenNC} for the NC decay mode. From Fig.~\ref{fig:SenCC} and 
Fig.~\ref{fig:SenNC}, we see that for $m_Q=400$ GeV, one would be able to reach a 5$\sigma$ discovery with 8 fb$^{-1}$ for 
$S_{\scriptscriptstyle D}^{\scriptscriptstyle CC} \ 
(S_{\scriptscriptstyle U}^{\scriptscriptstyle NC} ) \approx 0.2\ (0.4)$ and 
$S_{\scriptscriptstyle Q}^{\scriptscriptstyle CC(NC)} \approx 1$ for all other channels.  We tabulate the achievable sensitivity in $m_Q$ for 
$S_{\scriptscriptstyle Q}^{\scriptscriptstyle CC(NC)} = 1\ (2)$ in 
Table \ref{tab:mq}. In these figures, the current bound from direct $Q\bar Q$ searches at the Tevatron experiments is also indicated (the vertical lines).
Given the rather weak signal for the single top production at the Tevatron, we do not expect the recent observations for this channel \cite{Abazov:2006gd} to improve the sensitivity on $m_Q$ as presented in the figures.

\section{Summary}
We have presented a simple setup with vector-like doublets that satisfies all experimental constraints and can occur naturally in models with warped extra dimensions. In our set-up, the heavy quarks can have sizable gauge couplings to valence quarks. This large coupling along with enhanced parton luminosity and distinctive kinematics makes single production competitive with and in fact better than QCD pair production, especially for large masses. While we are motivated by such a scenario, we have performed our analysis and presented our results in a completely model-independent manner. We have found significant sensitivity in the parameter space of $m_Q$ and the model-dependent coupling and branching ratio of the heavy quarks. With 4 (8) fb$^{-1}$, one may reach a 
5$\sigma$ statistical significance for 580 (630) GeV for 
$S_{\scriptscriptstyle D}^{\scriptscriptstyle CC} = 1$ and 670 (710) GeV for 
$S_{\scriptscriptstyle D}^{\scriptscriptstyle CC} = 2$. According to Ref.~\cite{Mehdiyev:2007pf}, a $D$ quark with mass 640 (720) GeV and specific decay branching fractions can be discovered at 5$\sigma$ through pair production with $\sim 5\ (10$) fb$^{-1}$ data at the $\sqrt{s} = 14$ TeV LHC. There is also another study \cite{Sultansoy:2006rx} that estimates the LHC ($\sqrt{s} = 14$ TeV) reach of  heavy $D$ quarks via single quark production in the context of an $E_6$ model for a specific choice of parameters. A charge $2/3$ quark with mass 500 GeV decaying into third generation SM quarks can be discovered at 5$\sigma$ through pair production with $\sim 3 - 7$ fb$^{-1}$ data at the $\sqrt{s} = 14$ TeV LHC \cite{AguilarSaavedra:2005pv}. However, such bounds do not apply to our case where the quarks decay exclusively to first generation SM quarks. We have also suggested a method to effectively identify the electromagnetic charge of the produced heavy quark. In conclusion, we have investigated the potential to search for new heavy quarks at the Tevatron in a model-independent way and have found that the current sensitivity can be increased greatly in the context of the class of models presented by analysing the single quark production channel. Our study also shows that the Tevatron can probe an interesting class of extra-dimension models with this analysis.
\begin{table}[tb]
{
\begin{tabular}{ c |c c | c c }
\hline
$\int {\cal L}  dt$ & \multicolumn{2}{c}{4 fb$^{-1}$} &  \multicolumn{2}{c}{ 8 fb$^{-1}$} \\
\hline 
Sensitivity & 2$\sigma$  & 5$\sigma$ & 2$\sigma$  & 5$\sigma$ \\
\hline
$m_D$ for $S_{\scriptscriptstyle D}^{\scriptscriptstyle CC} = 1\ (2) $ & 720 (820) & 580 (670) & 760 (860) & 630 (710) \\
$m_U$ for $S_{\scriptscriptstyle U}^{\scriptscriptstyle CC} = 1\ (2) $ & 470 (530) & 370 (440) & 490 (560) & 400 (470) \\
$m_D$ for $S_{\scriptscriptstyle D}^{\scriptscriptstyle NC} = 1\ (2) $ & 450 (530) & 350 (420) & 490 (570) & 380 (470) \\
$m_U$ for $S_{\scriptscriptstyle U}^{\scriptscriptstyle NC} = 1\ (2) $ & 590 (680) & 460 (540) & 640 (730) & 510 (590) \\
\hline
\end{tabular}
}
\caption[]{Tevatron sensitivity for $m_{D, U}^{}$ (GeV).}
\label{tab:mq}
\end{table}

\vspace{1cm}

\begin{acknowledgments}
We acknowledge interesting discussions with C.~Csaki, J.~Hewett, E.~Ponton, T.~Rizzo, M.~Schmitt, C.~Wagner and D.~Zeppenfeld. We thank the hospitality of Kavli Institute of Theoretical Physics, Santa Barbara and the Aspen Center for Physics where part of this work was carried out. Fermilab is operated by Fermi Research Alliance, LLC under Contract No. DE-AC02-07CH11359 with the United States Department of Energy. The work of TH is supported in part by the United States Department of Energy under grant DE-FG02-95ER40896 and the Wisconsin Alumni Research Foundation. The work of JS is supported by SNSF grant 200021-117873. The work at KITP was supported in part by the National Science Foundation under Grant No. PHY05-51164.  
\end{acknowledgments}

\appendix
\section{Explicit Realization}

In this Appendix we describe in detail an explicit realization of a model with vector-like quarks that motivated the analysis in this paper. We will also comment on the features of five-dimensional models that make natural the apparent fine-tunings of four-dimensional models. The set-up is the SM extended with two vector-like quark $SU(2)_L$ doublets with hypercharges $1/6$ and $7/6$, denoted, respectively, by
\beq
Q^{(0)}_{L,R}=\left(\begin{array}{c} 
q^{(0)u}_{L,R} \\ q^{(0)d}_{L,R} \end{array}\right)_{1/6}, 
\quad
X^{(0)}_{L,R}=\left(\begin{array}{c} 
\chi^{(0)u}_{L,R} \\ \chi^{(0)d}_{L,R} \end{array}\right)_{7/6},
\eeq 
where we have denoted the hypercharge with a subscript. Note that the new quarks have electric charges equal to $2/3$ (for $q^u$ and $\chi^d$), $-1/3$ ($q^d$) and $5/3$ ($\chi^u$). We assume that these new vector-like quarks are exactly degenerate and couple (with identical strength) only to the up quark, in the basis in which all the SM flavor mixing occurs in the down sector. The Lagrangian reads in this basis
\bea
\nn
\mathcal{L} = \mathcal{L}_\mathrm{K}
&-&\Big[ \lambda^i_u \bar{q}^{(0)i}_L \tilde{\varphi} u^{(0)i}_R
+\lambda^j_d V_{ij} \bar{q}^{(0)i}_L \varphi d^{(0)j}_R \\
&+&\ \ \lambda_Q \big(\bar{Q}^{(0)}_L \tilde{\varphi}+\bar{X}^{(0)}_L \varphi\big)
u^{(0)}_R
+m_Q\big(\bar{Q}_L^{(0)} Q_R^{(0)} + \bar{X}_L^{(0)} X_R^{(0)}\big)
+ \mathrm{h.c.} \Big],\label{full:lag}
\eea
where $\mathcal{L}_\mathrm{K}=\bar{\psi} i \cancel{D} \psi$ is the sum of the diagonal kinetic terms (with covariant derivatives, thus including gauge couplings) for all the fields in the theory, $i,j=1,2,3$ are family indices, $V_{ij}$ is a unitary matrix (the CKM matrix in the absence of new physics), $\varphi$ is the SM Higgs field, $\tilde{\varphi}=i \sigma^2 \varphi^\ast $ and $u^{(0)}_R$ with no generational index is the SM up quark. All fields have a superscript 
$(0)$ to denote that they are not mass eigenstates. In order to extract the physics from this system, we need to go to the mass eigenstate basis, in which the mass Lagrangian is diagonal. Before doing that, we should notice that the charm and top quarks are already mass eigenstates. Furthermore, the charge $-1/3$ quark mass eigenstates are simply defined in terms of the current eigenstates as
\beq
d_L^i = V_{ij} d^{(0)j}_L, \quad d_R^i = d_R^{(0)i},
\quad q^d_{L,R}=q^{(0)d}_{L,R}.
\eeq 
Thus, the only non-trivial diagonalization comes from the mass Lagrangian involving the up quark (hereafter denoted simply by $u$, similarly from now on 
$\lambda_u \equiv \lambda_u^1$) and the charge $2/3$ quarks in the new vector-like multiplets. This diagonalization is done in two steps, first there is a redefinition of the heavy fields,
\beq
q^{(0)\pm}_{L,R}\equiv \frac{1}{\sqrt{2}}(q^{(0)u}\pm \chi^{(0)d}),
\eeq
so that the relevant part of the mass Lagrangian now reads,
\beq
\mathcal{L}=
\left( \begin{array}{ccc} 
\bar{u}^{(0)}_L & \bar{q}^{(0)+}_L & \bar{q}^{(0)-}_L
\end{array}\right)
\left( \begin{array}{ccc}
\lambda_u v & 0 & 0 \\
\sqrt{2} \lambda_Q v & m_Q & 0 \\
0 & 0 & m_Q 
\end{array}\right)
\left( \begin{array}{c}
u^{(0)}_R \\
q^{(0)+}_R \\
q^{(0)-}_R
\end{array}\right)
\eeq
This matrix can be diagonalized with the following two rotations,
\beq
\left( \begin{array}{c}
u^{(0)}_{L,R} \\
q^{(0)+}_{L,R} 
\end{array}\right)=
\left( \begin{array}{cc}
c_{L,R} & -s_{L,R} \\
s_{L,R} & c_{L,R} 
\end{array}\right)
\left( \begin{array}{c}
u_{L,R} \\
q^+_{L,R} 
\end{array}\right), \label{two:rotations}
\eeq
where $s_{L,R}\equiv \sin \theta_{L,R}$ and $c_{L,R}\equiv \cos \theta_{L,R}$. $q^-_{L,R}=q^{(0)-}_{L,R}$ is already a mass eigenstate with mass $m_Q$ and does not need to be rotated. The rotations in Eq.~(\ref{two:rotations}) can be computed exactly, but it is simpler to perturbatively expand the solution in the small parameter $v/m_Q$, as long as the relevant Yukawa couplings are at most order one, $|\lambda_u|,|\lambda_Q|\lesssim \mathcal{O}(1)$ and the new quarks are relatively heavy as compared with the Higgs vev, $v=174$ GeV, so that $v/m_Q\ll 1$. The result for the rotations is
\bea
\frac{s_L}{c_L}&=& - \sqrt{2} \lambda_u \lambda_Q
\left(\frac{v}{m_Q}\right)^2
\times \bigg[ 1 + (\lambda_u^2-2\lambda^2_Q)
\left(\frac{v}{m_Q}\right)^2
+\ldots \bigg], \\
\frac{s_R}{c_R}&=& - \sqrt{2} \lambda_Q
\frac{v}{m_Q}
\times \bigg[ 1 + \lambda_u\lambda_Q
\left(\frac{v}{m_Q}\right)^2
+\ldots \bigg], 
\eea
whereas the masses read,
\beq
m_u = \lambda_u v 
\bigg[ 1 - \lambda_Q^2 \left(\frac{v}{m_Q}\right)^2 +\ldots \bigg], 
\quad
m_{q^+} = m_Q 
\bigg[ 1 + \lambda_Q^2 \left(\frac{v}{m_Q}\right)^2 +\ldots \bigg].
\eeq
In particular we see that we need to take $\lambda_u\approx 10^{-5}$ in order to correctly reproduce the up quark mass. Note however that $\lambda_Q$ can be
order one without conflicting with the mass of the up quark. We will see below that an order one $\lambda_Q$ is in fact also compatible with electroweak and flavor constraints.

In the mass eigenstate basis, the gauge and Yukawa couplings are no longer
diagonal and get corrections with respect to their original values. The fact that we have introduced new fields with non-SM gauge quantum numbers, like right-handed $SU(2)_L$ doublets, prevents the GIM mechanism from protecting the gauge couplings in the physical basis. Similarly, the fact that the masses not only come from Yukawa couplings but also from direct Dirac masses lead to non-diagonal Yukawa interactions in the physical basis. We parametrize the new couplings in the physical basis as
\bea
\mathcal{L}^Z&=&- \frac{g}{2c_W} \Big[
\bar{\psi}^u_{iL} X^{L}_{ij} \gamma^\mu \psi^u_{jL}
-\bar{\psi}^d_{aL} X^{L}_{ab} \gamma^\mu \psi^d_{bL}
+ (L\to R) 
+\ldots
- 2 s_W^2 J^\mu_\mathrm{EM} \Big] Z_\mu,
\\
\mathcal{L}^W &=&
-\frac{g}{\sqrt{2}} \Big[
\bar{\psi}^u_{iL} W^L_{ia} \gamma^\mu \psi^d_{aL}
+\bar{\chi}^u_{L} \widetilde{W}^L_{\tilde{\chi}^uj}\gamma^\mu \psi^u_{jL}
+(L\to R)+\ldots
\Big]
W^+_\mu +\mathrm{h.c.},
\\
\mathcal{L}^H &=&-
\Big[
\bar{\psi}^u_{iL} Y^u_{ij} \psi^u_{jR}
+\bar{\psi}^d_{iL} Y^d_{ij} \psi^d_{jR}
+ \ldots+\mathrm{h.c.} \Big] \frac{H}{\sqrt{2}},
\eea
where $\psi^{u,d}$ represent all the quarks with charges $2/3$ and $-1/3$, respectively, the dots represent couplings that are not relevant for the phenomenology of these quarks and 
$J^\mu_\mathrm{EM} \equiv \sum_\psi \bar{\psi} \gamma^\mu Q \psi$ is the electromagnetic current. We separate these couplings in two categories: 
\begin{itemize}
\item The couplings of light quarks: These have been experimentally observed and are strongly constrained by electroweak precision and flavor data.
\item The couplings involving one light and one heavy quark: These have not been discovered yet and they give the main production and decay mechanism for the processes we have considered in this article. They can be constrained by flavor data if large flavor violations are introduced.
\end{itemize}  
Recall that the mass mixing only involves the up quark. Thus only the up and some of the down quark couplings can be modified, whereas the ones of the 
$s$, $c$, $b$ and $t$ quarks remain as in the SM. Regarding the first category of couplings, we obtain
\bea
X_{uu}^{R}&=&0,
\quad
W^R_{u d_i}=0, 
\quad
Y^u_{uu}=c_R(\lambda_u c_L+\sqrt{2} \lambda_Q s_L)
\approx \lambda_u \left(1-3\lambda_Q^2\frac{v^2}{m_Q^2}\right),
\nn \\
X_{uu}^{L}&=&c_L^2 \approx
1-2 \lambda_Q^2 \lambda^2_u \left(
\frac{v}{m_Q} \right)^4, 
\quad
W^L_{u d_i}=c_L\approx V_{ud_i}\left [ 1- \lambda^2_Q \lambda^2_u
 \left(\frac{v}{m_Q} \right)^4\right],\label{couplings:SM}
\eea
where the symbol $\approx$ indicates that we have neglected higher orders in the $v/m_Q$ expansion. We see that no new RH currents are introduced among the light quarks and that the LH gauge couplings get corrected at order 
$(v/m_Q)^4$. Furthermore, these corrections have an extra suppression proportional to $\lambda_u^2 \sim 10^{-10}$ and are therefore unobservable. Note that the correction to the up Yukawa coupling is still proportional to $\lambda_u$, with an additional suppression of $\lambda^2_Q (v/m_Q)^2$, and is therefore negligible. The relevant couplings between a light quark and a new heavy quark read
\beq
\left.
\begin{array}{ll} \displaystyle
X^{R}_{uq^-} = \sqrt{2}W^R_{u q^d} = \sqrt{2}\widetilde{W}^R_{\chi^u u} 
=s_R\approx -\sqrt{2} \lambda_Q \frac{v}{m_Q} \\
\displaystyle
\ Y^u_{q^+u} = c_R(-\lambda_u s_L + \sqrt{2} \lambda_Q c_L) 
\approx \sqrt{2}\lambda_Q \\
\end{array}
\right \}
\label{couplings:large}
\eeq
\beq
\left.
\begin{array}{ll} \displaystyle
\ \ \ \ \ \ \ \ \ X^{L}_{uq^-} = \sqrt{2}W^L_{u q^d} = \sqrt{2}\widetilde{W}^L_{\chi^u u} 
= s_L \approx -\sqrt{2} \lambda_u\lambda_Q \left(\frac{v}{m_Q}\right)^2 \\
\ \ \ \  \ \ \ \  \ X^{L}_{uq^+} = -s_L c_L \approx \sqrt{2} \lambda_u\lambda_Q \left(\frac{v}{m_Q}\right)^2 \\
\end{array}
\right \}
\label{couplings:small1}
\eeq
\bea
\label{couplings:fcnc}
\ W^L_{q^+d_i} &=&-V_{u d_i} s_L \approx
\sqrt{2} V_{u d_i} \lambda_u \lambda_Q \left(\frac{v}{m_Q}\right)^2,\\
\label{couplings:small2} 
\ Y^u_{uq^+}&=&-s_R(\lambda_u c_L+\sqrt{2} \lambda_Q s_L)
\approx \sqrt{2} \lambda_u \lambda_Q\frac{v}{m_Q}, \\
\label{couplings:zero}
\ X^R_{uq^+} &=& W^L_{q^-d_i} = W^R_{q^-d_i} = W^R_{q^+d_i} = Y^u_{uq^-} =
Y^u_{q^- u} = 0.
\eea
The couplings in Eq.~(\ref{couplings:large}) correspond to the ones we study in this article plus a non-suppressed Yukawa coupling that we do not investigate in this article. The gauge couplings are of order $v/m_Q$ and have no 
$\lambda_u$ suppression but only a coefficient $\lambda_Q$ or 
$\sqrt{2} \lambda_Q$, which can be order one and corresponds, in this approximation, to the relevant $\tilde{\kappa}$ parameter in the main text. Similarly, the last coupling is a Yukawa coupling that is of order $\lambda_Q$. It can lead to potentially interesting signatures in Higgs physics that will be investigated in a future publication. The couplings in
Eqs.~(\ref{couplings:small1}) and (\ref{couplings:fcnc}) are all of order 
$v^2/m_Q^2$ and have an extra $\lambda_u\sim 10^{-5}$ suppression. The coupling in Eq.~(\ref{couplings:fcnc}) in particular is the only new source of
flavor violation. The fact that it is suppressed by CKM angles and by 
$\lambda_u$ ensures no conflict with flavor data, even for $\lambda_Q \sim 1$. The coupling in Eq.~(\ref{couplings:small2}) is a Yukawa coupling between 
$q^+$  and the up quark, but it is suppressed by one power of $v/m_Q$ and most importantly by the tiny up Yukawa $\lambda_u\sim 10^{-5}$ and therefore has no phenomenological implications. Recall that there is another large Yukawa coupling between these two quarks that will be the main decay mode of $q^+$. Finally the couplings in Eq.~(\ref{couplings:zero}) are all identically zero. This analysis shows how one can have a scenario with large couplings of the up quark to new vector-like quarks without conflicting with current experimental data. A similar analysis could have been done for the down quark by adding two degenerate doublets with the right hypercharges ($1/6$ and 
$-5/6$). 

From the point of view of the four-dimensional model, however, there are a number of features that seem fine-tuned. In particular, we have introduced two exactly degenerate doublets with identical couplings to the SM quarks. We have assumed a large Yukawa coupling, $\lambda_Q\sim 1$, between the new quarks and the up quark, which itself has a tiny Yukawa coupling 
$\lambda_u\sim 10^{-5}$. Also, we have set to zero a Yukawa coupling between $Q_L^{(0)}$ and $d_R$, which is, \textit{a priori}, not forbidden by any symmetry. Finally, we have forbidden intergenerational mixing. However, as we said in the main text, all these features can appear naturally in models with extra
dimensions. For instance, degenerate bidoublets can appear naturally in models with custodial symmetry due to the fact that the two doublets are part of a gauge multiplet of the higher-dimensional symmetry. Another important property of the models with extra dimensions we are interested in is that not all Yukawa couplings that are allowed by the low energy (Standard Model) gauge
symmetries are actually allowed by the higher dimensional gauge symmetries. In particular there is a class of models in which there are only Yukawa couplings between a particular SM fermion and its ``own vector-like quarks'' (that correspond to its Kaluza-Klein excitations). This could explain why there is a Yukawa coupling with the up but not with the down quark, for instance. 

A final important point from a model-building perspective is the large difference in the size of the two Yukawas $\lambda_{u,Q}$, which can be easily understood from the structure of the Yukawas of the higher dimensional multiplets. Let us exemplify it in a simplified four-dimensional model that simulates the extra-dimensional set-up. Assume we have the matter content we have considered so far, the SM quarks plus two vector-like doublets with hypercharges $7/6$ and $1/6$, respectively, plus a vector-like singlet $U_{L,R}$ with hypercharge $2/3$ (same quantum numbers as the SM $u_R$ quark). Assume the two doublets and the $u_R$ live in the
same higher dimensional multiplet and therefore they have Yukawa couplings among themselves. Similarly, the new singlet is in the same multiplet as $q_L$ and therefore it only has Yukawa couplings with it. Finally, $u_R$ and $U_L$ have the same quantum numbers and therefore can have a direct Dirac mass coupling them (for simplicity, we will also assume that the allowed Dirac mass coupling $q_L$ and $q^u_R$ is vanishing). The resulting mass matrix Lagrangian reads  
\beq
\mathcal{L}=
\left( \begin{array}{cccc} 
\bar{u}^{(0)}_L & \bar{q}^{(0)u}_L & \bar{\chi}^{(0)d}_L & \bar{U}^{(0)}_L
\end{array}\right)
\left( \begin{array}{cccc}
0 & 0 & 0 &\tilde{\lambda}_u v\\
\tilde{\lambda}_Q v & m_Q & 0 & 0 \\
\tilde{\lambda}_Q v  & 0 & m_Q & 0 \\
m_1 & 0 & 0 & m_2
\end{array}\right)
\left( \begin{array}{c}
u^{(0)}_R \\
q^{(0)u}_R \\
\chi^{(0)d}_R \\
U^{(0)}_R
\end{array}\right) + \mathrm{h.c.}
\eeq
Note that this mass matrix is non-diagonal even in the absence of electroweak symmetry breaking (\textit{i.e.} $v=0$). In order to obtain a diagonal matrix in the absence of electroweak symmetry breaking, we perform the following rotation,
\beq
\left( \begin{array}{c}
u^{(0)}_{R} \\
U^{(0)}_{R} 
\end{array}\right) 
\to
\left( \begin{array}{cc}
c_{0} & -s_{0} \\
s_{0} & c_{0} 
\end{array}\right)
\left( \begin{array}{c}
u^{(0)}_{R} \\
U^{(0)}_{R} 
\end{array}\right), \label{two:rotation0}
\eeq
with $s_0/c_0\equiv \sin \theta_0/\cos \theta_0 = - m_1/m_2$. In that case the mass Lagrangian reads,
\beq
\mathcal{L}\to
\left( \begin{array}{cccc} 
\bar{u}^{(0)}_L & \bar{q}^{(0)u}_L & \bar{\chi}^{(0)d}_L & \bar{U}^{(0)}_L
\end{array}\right)
\left( \begin{array}{cccc}
s_0 \tilde{\lambda}_u v & 0 & 0 &c_0\tilde{\lambda}_u v\\
c_0\tilde{\lambda}_Q v & m_Q & 0 & -s_0\tilde{\lambda}_Q v  \\
c_0\tilde{\lambda}_Q v  & 0 & m_Q & -s_0\tilde{\lambda}_Q v  \\
0 & 0 & 0 & m_2/c_0
\end{array}\right)
\left( \begin{array}{c}
u^{(0)}_R \\
q^{(0)u}_R \\
\chi^{(0)d}_R \\
U^{(0)}_R
\end{array}\right) + \mathrm{h.c.}
\eeq
If $m_2/c_0$ is much larger than $m_Q$, the effects of the singlet can be safely neglected and we just have to look at the $3\times 3$ submatrix involving the SM quarks and the two doublets. In that case, we recover our original Lagrangian, Eq.~(\ref{full:lag}), with the identifications 
\beq
\lambda_u = s_0 \tilde{\lambda}_u, \quad
\lambda_Q = c_0 \tilde{\lambda}_Q.
\eeq
Thus, starting with all Yukawa couplings order one, 
$\tilde{\lambda}_u\sim \tilde{\lambda}_Q \sim 1$, we see that
$\lambda_u \sim 10^{-5}$ requires $s_0 \sim 10^{-5}$ and therefore
$\lambda_Q \approx \tilde{\lambda}_Q \sim 1$. Note that if the up quark was heavier (or for heavier generations), then it is not necessary that $s_0$ be this small, in which case $c_0$ would have been smaller resulting in the coupling 
$\lambda_Q$ being smaller as well. Thus, in this class of constructions which can naturally appear in models with extra dimensions, the coupling $\lambda_Q$ is large, \textit{precisely} because the up quark is so light.

An example of these features is the model in Ref.~\cite{Carena:2006bn}. This model was constructed with electroweak symmetry breaking in mind, which only involves third generation quarks. It has however all the ingredients to realize the
set-up we study in this paper. There are light Kaluza-Klein excitations of five-dimensional quarks (which are vector-like), with masses in the range $300-500$ GeV, that come in degenerate doublets with hypercharges $7/6$ and $1/6$. As we mentioned, they are degenerate because they belong to the same multiplet of a higher symmetry, a bidoublet of SO(4), and have the same boundary conditions. Furthermore, the two degenerate doublets and $q^{(0)}$ live in different five-dimensional multiplets, whereas $u^{(0)}_R$ lives in a superposition of both explaining that $\lambda_Q$ (\textit{i.e.} the corresponding $\tilde{\kappa}$) is large, precisely because the up quark is so light ($\lambda_u$ so small). Finally, although flavor issues were not discussed in that reference, the flavor constraints in warped models~\cite{flavorbounds}, require some sort of flavor symmetry protection~\cite{flavorprotection} that could easily explain why intergenerational mixing is forbidden. As an example, one
point for the model studied in Ref.~\cite{Carena:2006bn} gives 
$m_U=m_D=480$ GeV and $\tilde{\kappa}_{uD}=0.57$, 
$\tilde{\kappa}_{uU}=0.81$. The fact that all these ingredients can naturally appear in successful models of electroweak symmetry breaking has motivated us to perform this study.


\end{document}